\documentstyle[preprint,revtex]{aps}

\begin{document}
\draft
\begin{title}
Heavy Charged Leptons in an $SU(3)_L\otimes U(1)_N$ Model
\end{title}
\author{ V. Pleitez and M.D. Tonasse}
\begin{instit}
Instituto de F\'\i sica Te\'orica \\
Universidade Estadual Paulista \\
Rua Pamplona, 145 \\
01405-900--S\~ao Paulo, SP \\
Brazil
\end{instit}
\begin{abstract}
We consider an $SU(3)_L\otimes U(1)_N$ model for the electroweak
interactions which includes extra charged leptons which do
not mix with the known leptons. These new leptons couple to $Z^0$
only through vector currents. We consider constraints on the mass of
one of these leptons coming from the $Z^0$-width and from the
muon $(g-2)$ factor. The last one is less restrictive than the former.
\end{abstract}
\pacs{PACS numbers: 12.15.Cc; 14.60.Jj }
Recently, it was pointed out that models with $SU(3)_c \otimes
SU(3)_L \otimes U(1)_N$ symmetry have
interesting features and could be possible extensions of the standard
model for the interactions of quarks and leptons~\cite{pp,pp2,pp3}.

In order to make the model anomaly free, two of the three quark
generations transform identically and one
generation, it does not matter which, transforms in a different
representation of $SU(3)_L \otimes U(1)_N$. One can easily check that
all gauge anomalies cancel out in this theory.  Although each
generation is anomalous this type of construction is only anomaly
free when the number of generations is divisible by 3. Thus 3
generations is singled out as the simplest non-trivial anomaly free
$SU(3)_c \otimes SU(3)_L \otimes U(1)_N$ model.

Here we will consider a model which has the same quark sector of
Ref.~\cite{pp} but with a different leptonic
sector. Let us start by defining the charge operator as
\begin{equation}
{Q\over e}= \frac{1}{2}\left(\lambda_3 -\sqrt{3}\lambda_8\right)+N,
\label{q}
\end{equation}
the $\lambda$'s being the usual Gell-Mann matrices of $SU(3)$ and $N$
is the $U(1)_N$ charge.

The leptonic sector includes beside the usual charged leptons and
their respective neutrinos, charged heavy leptons and
transform as triplets $({\bf3},0)$ under $SU(3)_L\otimes U(1)_N$
\begin{equation}
\psi_{aL}=\left(\begin{array}{c}
\nu'_a \\ e'^-_a \\ E'^+_a\end{array}\right)_L\;\sim ({\bf3},0),
\label{l}
\end{equation}
where $a=1,2,3$ is the family index. The primed
fields are symmetry eigenstates. In the $SU(3)_L\otimes U(1)_N$ model
considered in Ref.~\cite{pp} all charged leptons degrees of freedom,
i.e., $e^a_L,(e_R)^c$ belong to the same triplet. Hence there is no
right-handed singlets. However, in the present model, only
left-handed fields appear in Eq.~(\ref{l}) and for this reason we have
to add the corresponding right-handed components,
$e'^-_{aR}\sim({\bf1},-1)$ and $E'^+_{aR}\sim({\bf1},+1)$. The
introduction of right-handed neutrinos is optional.

The addition of the right-handed singlets does not change the
anomaly cancellation since the new leptons have opposite charge
in relation to the known charged leptons.

The model has an extra global $U(1)$ symmetry. Hence we can assign a
lepton number for every field in Eq.~(\ref{l}).
Assuming that all fields in
Eq.~(\ref{l}) have the same lepton number, the mass
term $\bar\psi^c_{aL}\psi_{bL}$ is forbidden. Thus there is no mixing
between the $e'_{aL}$ and $E'_{aL}$ leptons. Since we have not
introduced right-handed neutrinos, they remain massless.

In order to generate lepton masses, we introduce the following
Higgs triplets $\rho$ and $\chi$:
\begin{equation}
\left(\begin{array}{c}
\rho^+ \\ \rho^0 \\ \rho^{++}
\end{array}\right) \;\sim ({\bf3},1);\qquad\qquad
\left(\begin{array}{c}
\chi^- \\ \chi^{--} \\ \chi^0\end{array}\right)\;\sim ({\bf3},-1).
\label{h}
\end{equation}
The third Higgs triplet $\eta\sim({\bf3},0)$ which is needed in the
quark sector would give a mass to the neutrinos if we have
introduced right-handed neutrinos transforming as singlets under
$SU(3)_L\otimes U(1)_N$. Here we will not consider in detail
the gauge sector since it is the same of Ref.~\cite{pp}. Let us just
remember that there are three charged vector bosons $W^+,V^+,U^{++}$
and two neutral ones $Z$ and $Z'$.

The Yukawa interactions in the leptonic sector is
\begin{equation}
{\cal
L}_l=-G_{ab}\bar\psi_{aL}e'^-_{bR}\rho-G'_{ab}\bar\psi_{aL}E'^+_{bR}\chi+H.c.
\label{y}
\end{equation}
In particular the mass term is
\begin{equation}
{\cal L}_m=-uG_{ab}\bar e'_{aL}e'_{bR}-wG'_{ab}\bar E'_{aL}E'_{bR}+H.c.,
\label{mt}
\end{equation}
where $u$ ($w$) is the vacuum expectation value (VEV) of the $\rho^0$
($\chi^0$). We can diagonalize the matrices in Eq.~(\ref{mt}) defining
\[e'^-_{aL}=U^L_{ab}e^-_{bL},\quad e'^-_{aR}=U^R_{ab}e^-_{bR},\]
\begin{equation}
E'^+_{aL}=V^L_{ab}E^+_{bL},\quad E'^+_{aR}=V^R_{ab}E^+_{bR},
\label{md}
\end{equation}
the unprimed fields are mass eigenstates which will be denoted by
$e^-,\mu^-,\tau^-$ and $E^+,M^+,T^+$.  Notice that the masses of the
new leptons are proportional to the VEV, $w$, which is in control
of the $SU(3)$ symmetry and could be very heavy.

Next, let us consider the current-vector boson interactions which are
read off from
\begin{equation}
{\cal L}_F=\bar Ri\gamma^\mu(\partial_\mu+ig'B_\mu N_R)R+
\bar Li\gamma^\mu(\partial_\mu+\frac{ig'}{2}B_\mu
N_L+\frac{ig}{2}\vec\lambda\cdot \vec W_\mu)L,
\label{int}
\end{equation}
where $R$ ($L$) denotes any right-handed singlet (left-handed
triplet).

The electric charge is defined as
\begin{equation}
\vert e\vert=\frac{g\sin\theta}{(1+3\sin^2\theta)^{\frac{1}{2}}},
\label{ec}
\end{equation}
with $t\equiv\tan\theta= g/g'$. In terms of the weak mixing angle $\theta_W$
defined as $1/\cos\theta_W=M_Z/M_W$, we can write down
\begin{equation}
t^2= \frac{\sin^2\theta_W}{1-4\sin^2\theta_W}.
\label{1}
\end{equation}
The charged current interactions in terms of the mass eigenstates are
\begin{eqnarray}
{\cal L}_{CC}&=&-\frac{g}{\sqrt2}\sum_a[\bar
e_{aL}\gamma^\mu\nu_{aL}W^-_\mu+\bar
E_{aL}\gamma^\mu K_{ab}\nu_{bL}V^+_\mu \nonumber \\
\mbox{} & & +\bar e_{aL}\gamma^\mu
K^\dagger_{ab}E_{bL}U^{--}_\mu]+H.c.,
\label{cc}
\end{eqnarray}
where we have defined $\nu_{aL}=U^{L\dagger}_{ab}\nu'_{bL}$ and the
mixing matrix $K$ is defined as $K=V^{L\dagger}U^L$.
The gauge boson $V^+_\mu$ and $U^{--}_\mu$,
have a mass greater than $4$ TeV~\cite{pp}.

In this model there is also an additional neutral vector boson,
$Z'^0$, with a mass greater than $40$ TeV~\cite{pp}.
The neutral currents have the form
\begin{equation}
J^\mu_Z=-\frac{g}{2c_W}[a_{L}(f)\bar f\gamma^\mu(1-\gamma_5)f+a_{R}(f)\bar
f\gamma^\mu(1+\gamma_5)f],
\label{ncz}
\end{equation}
coupled to the $Z^0$, and
\begin{equation}
J^\mu_{Z'}=-\frac{g}{2c_W}[a'_{L}(f)\bar f\gamma^\mu(1-\gamma_5)f+
a'_{R}(f)\bar f\gamma^\mu(1+\gamma_5)f],
\label{ncr2}
\end{equation}
coupled to the $Z'^0$, where $f$ denotes any fermion.

Explicitly one has
\[
a_L(\nu'_a)=\frac{1}{2}, \quad a_R(\nu_a)=0,
\]
\begin{equation}
a'_L(\nu'_a)=\frac{1}{2}\left( \frac{1-4x}{3}
\right)^{\frac{1}{2}},\quad a'_R(\nu'_a)=0,
\label{a}
\end{equation}
for neutrinos,
\[
a_L(e'_a)=-\frac{1}{2}+x,\quad a_R(e'_a)=x,\]
\begin{equation}
a'_L(e'_a)=\frac{1}{2}\left( \frac{1-4x}{3}
\right)^{\frac{1}{2}},\quad a'_R(e'_a)=-x\left
(\frac{3}{1-4x}\right)^{\frac{1}{2}},
\label{a2}
\end{equation}
for the lightest charged leptons $e_a$, and
\[a_L(E'_a)=a_R(E'_a)=-x,\]
\begin{equation}
a'_L(E'_a)=-\left(\frac{1-4x}{3}\right)^{\frac{1}{2}},\quad
a'_R(E'_a)=x\left(\frac{3}{1-4x}\right)^{\frac{1}{2}},
\label{a3}
\end{equation}
for the heavy charged leptons $E_a$, with $x\equiv\sin^2\theta_W$.

Notice that the heavy leptons $E'_a$ have pure vector current
interactions with the $Z^0$ neutral gauge boson. From the
Eqs.~(\ref{a})--(\ref{a3}) we see that the GIM mechanism is
implemented.

The charged Higgs boson interaction Lagrangian is
\begin{eqnarray}
{\cal
L}_{Yl}&=&-\frac{m_a}{u}\bar\nu_{aL}e_{aR}\rho^+-\frac{m_b}{u}K_{ab}\bar
E_{aL}e_{bR}\rho^{++}\nonumber \\ \mbox{} &
&-\frac{M_b}{w}K_{ab}\bar\nu_{aL}E_{bR}\chi^--
\frac{M_b}{w}K^\dagger_{ab}\bar e_{aL}E_{bR}\chi^{--}+H.c.
\label{higgs}
\end{eqnarray}
where $m_a$ and $M_a$ denote the mass of $e_a$ and $E_a$ leptons
respectively. The heavy charged leptons
in this model do not belong to any of the
four types of heavy leptons usually considered in the literature: i)
sequential leptons, ii) paraleptons, iii) ortholeptons, or iv)
 long-lived penetrating particles, hence the experimental limits
already existing~\cite{pdg} do not apply directly to them.

For example, direct search for heavy leptons produced in $e^+e^-$
colliders have been
done, but with the heavy leptons being assumed coupled to the $Z^0$ in
the same way as ordinary leptons. These
heavy sequential charged and neutral leptons have been excluded
except if both masses are larger than $42.8$ GeV~\cite{l3}. In the
present model the new charged leptons have couplings to the neutral
vector bosons $(Z^0,Z'^0)$ which are different from those of the
lightest leptons as can be see from Eqs.~(\ref{a2}) and (\ref{a3}).

The main decay modes of the new leptons are those among the exotic
leptons themselves as $M^+\to E^+\nu_e\bar\nu_\mu$, assuming
$M_M>M_E$.

Search for these charged leptons could be made in colliding-beam
experiments $e^+e^-\to L^+L^-$, pair photoproduction
$\gamma+Z\to L^+L^-Z^*$ and in neutrino experiments
$\nu+Z\to L^-+Z^*$ where $L^\pm$ denotes $E^\pm,M^\pm$ or $T^\pm$~\cite{t}.

Decays as $L^+\to\bar\nu_L+hadrons$, imply
hadrons involving exotic quarks with charge $5/3$ and $-4/3$.

Next, we consider some possible constraints on the new leptons masses.
The partial decay width for the $Z^0$ decaying into
massless fermions is~\cite{pdg} $\Gamma(Z\to \bar
ff)=C(G_FM^3_Z)/6\sqrt{2}\pi)[V_i^2+A_i^2]$,
where $V^i=a^i_L+a^i_R$ and $A^i=a^i_L-a^i_R$ with $a^i_L, a^i_R$
defined in Eqs.~(\ref{a})--(\ref{a3}). For leptons $C=1$. We see that
$V_{e_a}=-1/2+2x$,
$A_{e_a}=-1/2$ and $V_{E_a}=-2x$, $A_{E_a}=0$. In the massless
limit the contribution of one of the exotic lepton, say $E^+$, to the
$Z^0$ partial width is $0.84\times\Gamma(Z\to e^+e^-)$. It means that
we have to consider corrections for massive fermions. Assuming that
$M_E<M_Z/2$ one has~\cite{new}
\begin{equation}
\Gamma(Z\to \bar
EE)=\frac{G_FM^3_Z}{6\sqrt{2}\pi}(1-4\mu_E)^{\frac{1}{2}}(1+2\mu_E)V_E^2,
\label{width}
\end{equation}
since $A_E=0$ and $\mu_E=M^2_E/M_Z^2$. Assuming that the contribution
for the $Z\to\bar EE$ is of the order of $20$ MeV, which is compatible with
the uncertainty in the measurement of the total $Z^0$ width~\cite{delphi},
we conclude that the mass of the
$E^+$ must be restricted to the interval $38.17$--$45.57$ GeV. The
other two exotic leptons may have masses larger than $M_Z/2$.

On the other hand, there are contributions to muon $g-2$
via transitions like $\mu^-\to M^+U^{--}\to \mu^-$. As this quantity
has been measured very accurately~\cite{exp} it can be used to
restrict the range of the parameters in a given model for the
electroweak interactions~\cite{g}. General expressions valid for an
arbitrary gauge model have been given in Ref.~\cite{yuca}, from which
it is easy to verify that the contributions of a lepton-$E_a$ to
$a_\mu=(g-2)_\mu/2$ is
\begin{equation}
a_\mu^U=a_\mu^W\left(\frac{M_a}{m_\mu}\right)\times
\left(\frac{M^2_W}{M^2_U}\right)\sim
a_\mu^W\left(\frac{M_a}{m_\mu}\right)\times10^{-4},
\label{muon1}
\end{equation}
if $m_\mu\ll M_a\ll M_U$, where $a^W_\mu$ is the usual one-loop
weak contributions and $M_U$ is the $U^{--}_\mu$ vector boson mass.
Hence, these contributions are negligible even when $M_a\sim M_Z/2$.

There are also contributions to the muon $g-2$ due to charged scalars
as $\mu^-\to \chi^-\nu,\rho^-\nu\to \mu^-$ and
$\mu^-\to \chi^{--}M^+,\rho^{--}M^+\to \mu^-$. In this case a factor
$(M_a/m_\mu)(M^2_W/m^2_\phi)$ appears, where $\phi$ is any charged
scalar. These contributions are also negligible if $m_\phi\gg M_W$,
as is the case, since the charged scalars have masses proportional to
$w\simeq 8\, \mbox{TeV}$ which is the VEV that induce the first
symmetry breaking~\cite{pp}.

Anyway, the scalar
contributions could be negligible because the $\mu-\phi$ coupling is
small. There could be also a cancellation between
both contributions from charged vector and scalar bosons~\cite{h}.
Hence the muon $g-2$ is not so restrictive on the heavy lepton and
charged scalar masses. The contributions to electron $g-2$ are
smaller than those to the muon for factors $m_e/m_\mu$ or
$m^2_e/m_\mu^2$.
\acknowledgments
We would like to thank the
Con\-se\-lho Na\-cio\-nal de De\-sen\-vol\-vi\-men\-to Cien\-t\'\i
\-fi\-co e Tec\-no\-l\'o\-gi\-co (CNPq) for full (MDT) and partial
(VP) financial support.


\begin{references}
\bibitem{pp} F. Pisano and V. Pleitez, Phys. Rev. D{\bf 46},
410(1992); P. Frampton, Phys. Rev. Lett. {\bf 69}, 2889(1992).
\bibitem{pp2}  R.\ Foot, O.\ F.\ Hernandez, F.\ Pisano and V.\
Pleitez, Leptons masses in an $SU(3)_L\otimes U(1)_N$ Gauge Model,
to appear in Phys. Rev. D.
\bibitem{pp3} J.C. Montero, F. Pisano and V. Pleitez, Neutral
currents and GIM mechanism in $SU(3)_L\otimes U(1)_N$ models
for electroweak interactions, accepted for
publication in Phys. Rev. D.
\bibitem{pdg} Particle Data Group, Phys.\ Rev.\ D {\bf45}, Part
II(1992).
\bibitem{l3} B. Adeva et al. (L3 Collaboration), Phys. Lett.
B{\bf251}, 321(1990); M.Z. Akrawy et al., (OPAL Collaboration), Phys.
Lett. B{\bf240}, 250(1990).
\bibitem{t} Y. S. Tsai, Phys. Rev. D{\bf4}, 2821(1971); erratum:
Phys. Rev. D{\bf13}, 771(1976).
\bibitem{new} A. Sirlin, Phys. Rev. D{\bf22}, 971(1980); D. Albert,
W.J. Marciano, D. Wyler and Z. Parsa, Nucl. Phys. B{\bf166},
460(1980); W. Hollik, Fortsch. Phys. {\bf38}, 165(1990).
\bibitem{delphi} P. Abreu et al., (DELPHI Collaboration), Nucl. Phys. {\bf
B367}, 511(1991).
\bibitem{exp} J. Bailey et al., Phys. Lett {\bf68B}, 191(1977); F.J.
Farley and E. Picasso, Ann. Rev. Nucl. Sci. {\bf29}, 243(1979).
\bibitem{g}J. R. Primack and H. R. Quinn, Phys. Rev. D{\bf6},
3171(1972); J. D. Bjorken and C. H. Llewellyn Smith, Phys. Rev.
D{\bf7}, 887(1973); I. Bars and M. Yoshimura, Phys. Rev. D{\bf6};
374(1978); K. Fujikawa, B.W. Lee and A.I. Sanda, Phys. Rev. D{\bf6},
2923(1972); W.A. Bardeen, R. Gastmans and B.E. Lautrup, Nucl. Phys.
B{\bf46}, 3199(1972).
\bibitem{yuca} J. P. Leveille, Nucl. Phys. B{\bf137}, 63(1978); S. R.
Moore, K. Whistnat and B-L. Young, Phys. Rev. D{\bf31}, 105(1985).
\bibitem{h} R. Jackiw and S. Weinberg, Phys. Rev. D{\bf5},
2396(1972).
\end{references}
\end{document}